\documentstyle[pra,aps]{revtex}

\newcommand{\vecbo}[1]{\mbox{\boldmath $#1$}}

\def\ni{\noindent}

\def\pa {\partial}
\def\/{\over}
\def\meio{{1\over{2}}}
\def\lco {\left[}
\def\rco {\right]}
\def\lp {\left(}
\def\rp {\right)}
\def\lch {\left\{}
\def\rch {\right\}}

\def\z{\vecbo{\hat{z}}}

\def\k {\it \vecbo{k} }
\def\kpa {\it \vecbo{k}_{\|} }

\def\r {\it \vecbo{r} }
\def\rpa {\it \vecbo{r}_{\|} }

\def\dud {\it \vecbo{d}_{12}}

\def\dudpa {\it \vecbo{d}_{12}^{\|}}
\def\dudpe {\it \vecbo{d}_{12}^{\bot}}

\def\A {\vecbo{A}}

\def\Ape {A^{\bot}_{21}}
\def\Apa {A^{\|}_{21}}

\def\Afpe {A^{0\bot}_{21}}
\def\Afpa {A^{0\|}_{21}}

\def\al {\alpha }

\def\de {\delta }
\def\om {\omega }
\def\la {\lambda }

\begin{document}
\draft

\twocolumn[\hsize\textwidth\columnwidth\hsize\csname
@twocolumnfalse\endcsname

\title{Spontaneous emission between an unusual pair of plates}

\author{D. T. Alves$^{(1,2)}$\cite{email1}, C. Farina$^{(3)}$\cite{email2}
and A. C.
Tort$^{(3)}$\cite{email3}}

\address{${(1)}$ - Centro Brasileiro de Pesquisas F\'{\i}sicas, DCP,
Rua Dr. Xavier Sigaud, 150, RJ 22290-180, Brazil}

\address{${(2)}$ - Universidade Federal do Par\'a, Departamento de
F\'{\i}sica, Bel\'em,  PA 66075-000, Brazil}

\address{${(3)}$ - Instituto de F\'{\i}sica,
UFRJ, Caixa Postal 68528,
21945-970 Rio de Janeiro RJ, Brazil}

\date{\today}
\maketitle

\begin{abstract}
We compute the modification in the spontaneous emission rate for a
two-level atom when it is located between two parallel plates of
different nature: a perfectly conducting plate
$(\epsilon\rightarrow \infty)$ and an infinitely permeable one
$(\mu\rightarrow \infty)$. We also discuss the case of two
infinitely permeable plates. We compare our results with those found in the 
literature for
the case of two perfectly conducting plates.
\end{abstract}

 \vskip2pc]
\label{sec:level1}

\pagebreak

Using thermodynamic arguments and assuming that thermal
equilibrium between matter and radiation is allways achieved,
Einstein \cite{Einstein} was able to demonstrate that, besides
stimulated emisson, excited atoms must also decay spontaneously.
Even an \lq\lq isolated" excited atom in vacuum must inevitably
decay to the ground state. In other words, an excited stationary
state of an atom is not actually a stationary state and we can say
that spontaneous emission is in fact not a property of an isolated
atom, but of an atom-vacuum system \cite{HarocheKleppner}. In the
context of QED, we can say that the ultimate reason for
spontaneous emission of excited atoms is the interaction of the
atom with the quantized electromagnetic field of the vacuum state.
As a consequence, any modification in the vacuum electromagnetic
field, caused for instance by cavities, can modify in principle
the radiative properties of atomic systems. We can say that the
presence of material walls in the vicinity of atomic systems
renormalizes their transition frequencies as well as the widths of
their spectral lines. The former effect corresponds to the
influence of boundary conditions (BC) in the analogue of the Lamb
shift, while the latter corresponds to the change in the
spontaneous emission rate of excited atoms. The branch of physics
that is concerned with the influence of the environment of an
atomic system in its radiative properties is called generically
Cavity QED  and the above examples are only two among many others
(for a review see for instance ref(s)\cite{Haroche,Berman}. Here
 we shall be concerned with one of the above effects,
namely, the influence of BC imposed on the radiation field in the
spontaneous emission rate of a two-level atom. It is worth
mentioning that Cavity QED was born precisely by the observation
of Purcell \cite{Purcell} half a century ago that spontaneous
emission process associated with nuclear magnetic moment
transitions at radio frequencies could be enhanced if the system
were coupled to a ressonant external electric circuit placed in
the vicinity of the system. However, we can say that the first
detailed papers on this subject were those written by Casimir and
Polder \cite{Polder} in which, among other things, forces between
polarizable atoms and metallic walls were treated, and by Casimir
in his seminal work that brought about the Casimir effect \cite{Casimir}. 
Since then, Cavity QED has
attracted the attention of many physicists, both theoretical and
experimentalists. Particularly, the effects of the proximity of
plane walls to atomic systems have been investigated: for
instance, Morawitz \cite{Morawitz} discussed both classically and
quantum-mechanically the influence of a plane mirror in the
spontaneous emission rate of a two-level atom. A few years later,
Drexhage \cite{Drexhage70/74} observed experimentally the
oscillatory behaviour of the lifetime on the distance to the
mirror. The QED of charged particles between two parallel mirrors
was discussed extensively by Barton \cite{Barton70,Barton87}, who
was the first to compute explicitly the influence of two parallel
perfect conducting plates in the spontaneous emission rate for a
spherically averaged atomic transition \cite{Barton70}. Barton's
result was rederived by Philpott \cite{Philpott73} with a similar
method and by Milonni and Knight \cite{MilonniKnight73} in the
context of the image method. An interesting feature of the
modified spontaneous emission rate between two conducting mirrors is the fact
that for the case of a transition dipole moment parallel to the
plates there must be a strong suppression for $2L/\lambda<1$, where
$L$ is the distance between the plates and $\lambda$, the
transition wavelength (see for instance ref.\cite{Milonnibook}).
This inhibited spontaneous emission has been observed
experimentally by Hulet, Hilfer and Kleppner
\cite{HuletHilferKleppner85}. Many other interesting experiments
have been done and we suggest for the interested reader the
reviews by Haroche and Kleppner \cite{HarocheKleppner} and Hinds
\cite{Hinds90} and references therein.

\bigskip
In this letter we compute the spontaneous emission
rate for a two-level atom when it is located between two parallel
plates of different nature ($\epsilon\rightarrow \infty$ and
$\mu\rightarrow \infty$) and between two infinitely permeable
plates ($\mu\rightarrow \infty$), and then, we compare our results with
those found in the literature \cite{Barton70,Philpott73,MilonniKnight73} for
the case of two perfectly conducting plates. Though analogous, the results
are different, since when we change the boundary conditions on the photon
field, the vacuum field modes also change. As expected, a strong suppression
also occurs for both cases treated here. However, curious as it may seem,
this suppression occurs when the transition dipole moment is perpendicular
to the plates, in contrast to the suppression when the dipole moment is
parallel to the plates that occurs for the two perfectly conducting plates.

\bigskip
Our starting point is the general expression for the spontaneous emission
rate of a transition 
$ 2 \rightarrow 1 $ of a two-level atom, which is given by:

\begin{equation}
A_{21}(\r) = \frac{4
{\pi}^{2}{\omega_{0}}^2}{\hbar}\sum_{\alpha}{1\over{\omega_{\alpha}}}
|\A_{\al}(\r)\cdot \dud |^{2} \de (\om_{\al}-\om_{0}), \label{m6.5}
\end{equation}

\ni where $\om_{0}$ corresponds to the transition frequency,
 $\dud$ is the transition dipole moment and each mode $\A_{\al}(\r)$ of the
vacuum field is characterized by a wave vector
$\k$ and a polarization $\la$ (see for instance ref.\cite{Milonnibook}).

The first setup we will consider consists of two infinite parallel
surfaces (the plates) one of which will be considered to be a
perfect conductor ($\epsilon\to\infty$) while the other is
supposed to be perfectly permeable ($\mu\to\infty$). Also, we will
choose Cartesian axes in such a way that the axis ${\cal OZ}$ is
perpendicular to both surafces. The perfectly conducting surface
will be placed at $z=0$ and the permeable one, at $z=L$. The
electromagnetic fields must satisfy the following boundary
conditions: {\bf (a)} the tangential components $E_x$ and $E_y$ of
the electric field as well as the normal component $B_z$ of the
magnetic field must vanish on the metallic plate at $z=0$. ({\bf
b)} The tangential components $B_x$ and $B_y$ of the magnetic
field must vanish on the permeable plate at $z=L$. It is
convenient to work with the vector potential
$\vecbo{A}(\vecbo{r},t)$ in the Coulomb gauge in which
$\vecbo{\nabla\cdot A}(\vecbo{r},t)=0$,
$\vecbo{E}(\vecbo{r},t)=-\partial\vecbo{A}(\vecbo{r},t)/\partial
t$ and
$\vecbo{B}(\vecbo{r},t)=\vecbo{\nabla\times}\vecbo{{A}(\vecbo{r},t)}$.
With this choice of gauge, the above boundary conditions can be  written as
conditions 
imposed on the vector potential components:
\ni

\begin{equation}
 A_{x}(x,y,0) = A_{y}(x,y,0)={\pa A_{z}\over{\pa z}}(x,y,0)
=0 
\end{equation}
\begin{equation}
 {\pa A_{x}\over{\pa z}}(x,y,L) ={\pa A_{y}\over{\pa z}}(x,y,L)=
A_{z}(x,y,L)= 0 
\end{equation}
\label{boundary1}

\ni The mode functions for this case are \cite{Bartao}:

\begin{equation}
\A_{\k 1}(\r) =
{\lp{2\over{V}}\rp}^{1/2}(\kpa\times\z)\sin{\lco{( n +
\meio)}{{\pi z}\/L} \rco e^{i\kpa\cdot\rpa}}
\end{equation}
and
\begin{equation}
\begin{array}{l}
  \A_{\k 2}(\r) = {\lp 1\/k \rp}
{\lp{2\over{V}}\rp}^{1/2}e^{i\kpa\cdot\rpa}\times\\\\
 \times{\lch {\it {k}_{\|} }
 \z
\cos{\lco{( n + \meio)} {{\pi z}\/L} \rco} -i {\pi\/{L}}{({{n + \meio}
}) \kpa \sin{\lco{( n + \meio)}{{\pi z}\/L} \rco} } \rch}
\end{array}
\end{equation}

\ni The contributions for the spontaneous emission rate associated with
$\dudpe$ and $\dudpa$ are given respectively by:

\begin{equation}
\Ape(z) = {3\pi \/{k_{0}L}}\Afpe \sum_{n=0}^{N} \lco 1 - {{\lp
n+\meio\rp}^{2}\/{k_{0}^{2} L^{2}}}\pi^{2} \rco \cos^{2}{ \lco{( n
+ \meio)}{{\pi z}\/L} \rco }
\end{equation}
and
\begin{equation}
\Apa(z) = {3\pi \/ 2{k_{0}L}}\Afpa \sum_{n=0}^{N} \lco 1  + {{\lp
n+\meio\rp}^{2}\/{k_{0}^{2} L^{2}}}\pi^{2} \rco \sin^{2}{ \lco{( n
+ \meio)}{{\pi z}\/L} \rco }
\end{equation}

\ni where $\Afpe$ and $\Afpa$are the corresponding contributions for the
spontaneous emission 
rate in unbounded (free) space, namely:

\begin{equation}
\Afpa = {4 |\dudpa|^{2}\om_{0}^{3}\/{3\hbar c^{3}}}
\;\;\;\mbox{and}\;\;\;
\Afpe = {4 |\dudpe|^{2}\om_{0}^{3}\/{3\hbar c^{3}}}
\end{equation}

\ni and $N$ is the greatest integer part of $k_{0} L/\pi-1/2$. The total
emission coefficient is given by ${A_{21}=\Ape +
\Apa}$ Recall that Einstein's coefficient for spontaneous emission is simply
given by
\begin{equation}
A^0_{21} = \Afpa + 
\Afpe={4 |\dud|^{2}\om_{0}^{3}\/{3\hbar c^{3}}}
\end{equation}

The graph displayed in figure (\ref{1}) shows the ratio between $A_{21}$ and
$A_{21}^0$ as a function of the dimensionless variable $s:=k_0z$ for the
case of two conducting plates (dashed line) and the case of a conducting
plate and a permeable plate (solid line). Although the two curves are
analogous, in the sense that both present oscillations with $s$, they are
different curves since the mode functions of the vacuum field in each case
are not the same. It is worth emphasizing the lack of symmetry of the latter
curve around the point that is equidistant from the plates. This was
expected because in this case the two plates correspond to distinct
electromagnetic media, with different properties.

The second example we shall be concerned with consists of two perfectly
permeable plates. The boundary conditions for this case can be cast into the
form:
\begin{equation}\label{boundary2a}
{\pa A_{x}\over{\pa z}}(x,y,0) ={\pa A_{y}\over{\pa z}}(x,y,0)=
A_{z}(x,y,0)= 0 
\end{equation}
\begin{equation}\label{boundary2b}
{\pa A_{x}\over{\pa z}}(x,y,L) ={\pa A_{y}\over{\pa z}}(x,y,L)=
A_{z}(x,y,L)= 0 
\end{equation}

\ni The corresponding mode functions are:

\begin{equation}
\A_{\k 1}(\r) =
{\lp{2\over{V}}\rp}^{1/2}(\kpa\times\z)\cos{({n}{{\pi z}\/L})
e^{i\kpa\cdot\rpa}}
\end{equation}
and
\begin{equation}
\begin{array}{l}
\A_{\k 2}(\r) = {\lp 1\/k \rp}
{\lp{2\over{V}}\rp}^{1/2}e^{i\kpa\cdot\rpa}\times\\\\
\times{\lch {\it {k}_{\|} }
\z
\sin{({n } {{\pi z}\/L} )} -i{( {{n \pi}
\/{L}}) \kpa \cos{({n } {{\pi z}\/L} )} } \rch}
\end{array}
\end{equation}

\ni The contributions for the spontaneous emission rate associated with
$\dudpe$ and $\dudpa$ are given respectively by:

\begin{equation}
\Ape(z) = {3\pi \/{k_{0}L}}\Afpe \sum_{n=1}^{N} \lp 1 - {{
n }^{2}\/{k_{0}^{2} L^{2}}}\pi^{2} \rp \sin^{2}{({ n}{{\pi z}\/L} ) }
\end{equation}
and
\begin{equation}
\Apa(z) = {3\pi \/{ 2k_{0}L}}\Afpa \lch \meio + \sum_{n=1}^{N} \lp 1  + {{
n}^{2}\/{k_{0}^{2} L^{2}}}\pi^{2} \rp \cos^{2}{ ({n}{{\pi z}\/L} ) } \rch
\end{equation}

Figure (\ref{2}) shows the ratio between $A_{21}$ and $A_{21}^0$ as a
function of $s=k_0z$ for the case of two conducting plates (dashed line) and
the case of two permeable plates (solid line). The curve for this latter
case also presents oscillations in the spontaneous emission rate as the
distance from the atom to each plate varies and is also symmetric with
respect to the equidistant point to the plates. However, there is a
remarkable difference between these two curves: whenever there is an
enhancement in the spontaneous emission rate of the former, there  will be a
depletion for the latter and vice versa. Particularly, their behaviour near
the plates are quite different.

\begin{figure}
\caption{The ratio $\Ape/\Afpe$ for the case of one perfectly conducting and
one infinitely permeable plate. Suppression occurs at $l=k_0L=\pi/2$.
}
\label{4}
\end{figure}
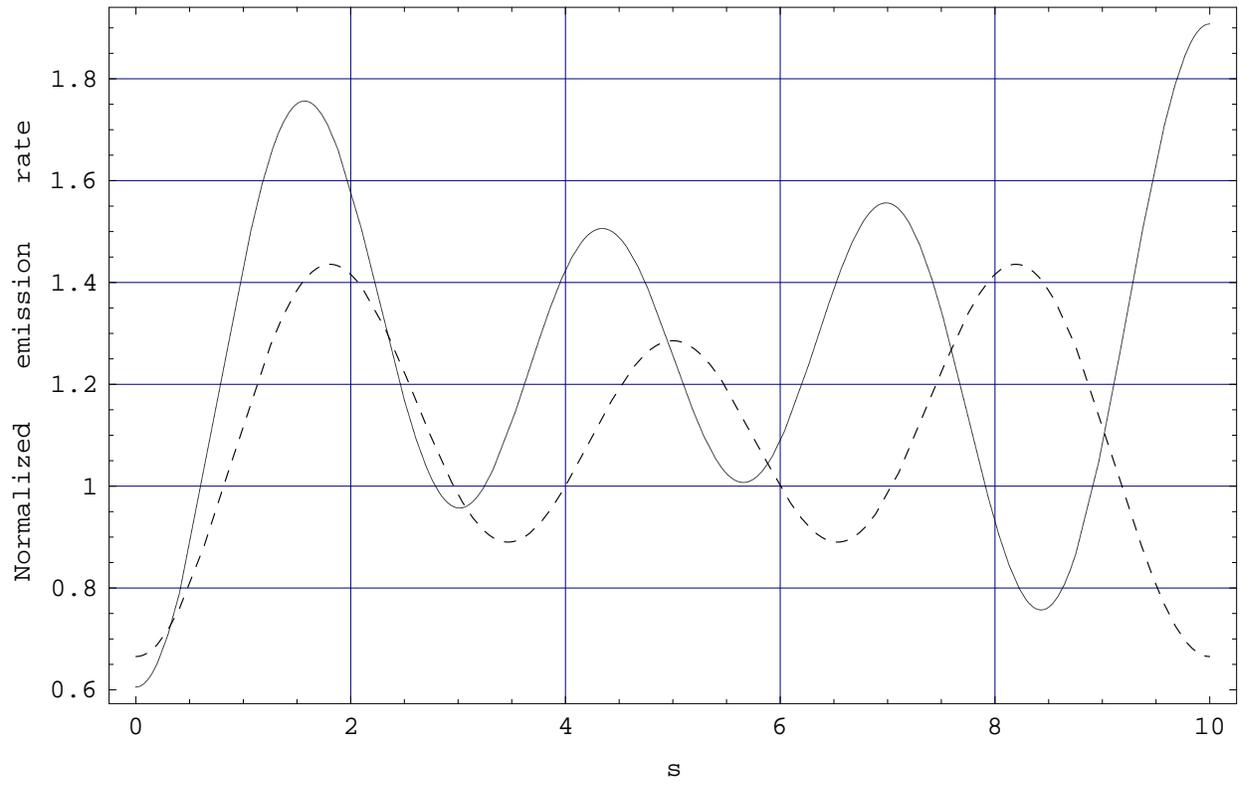
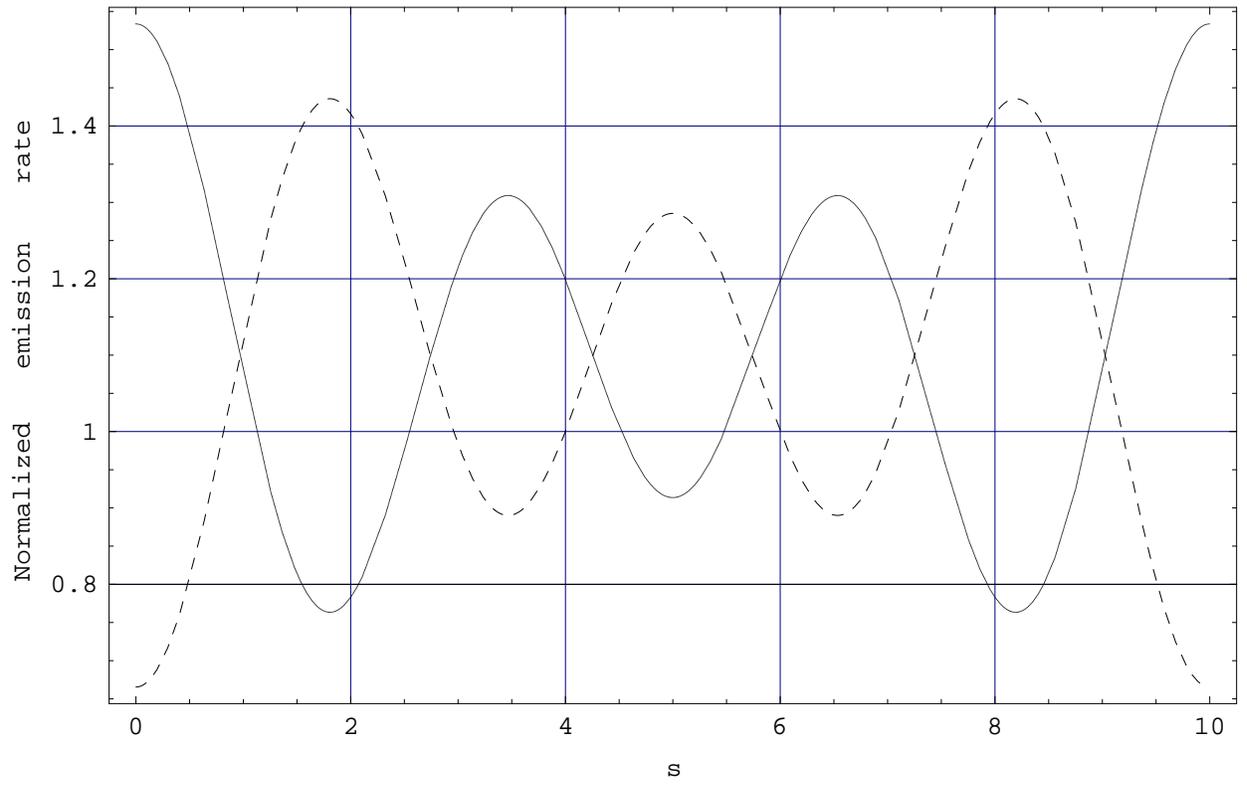
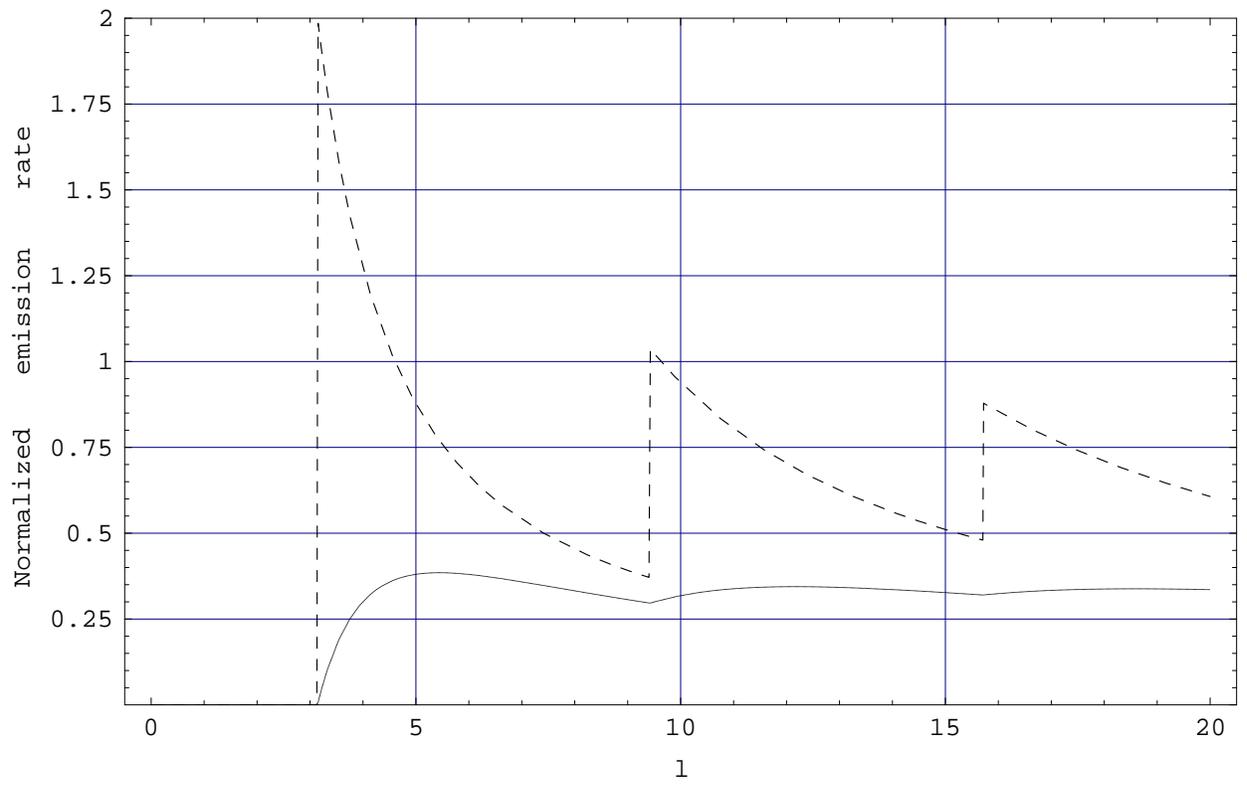
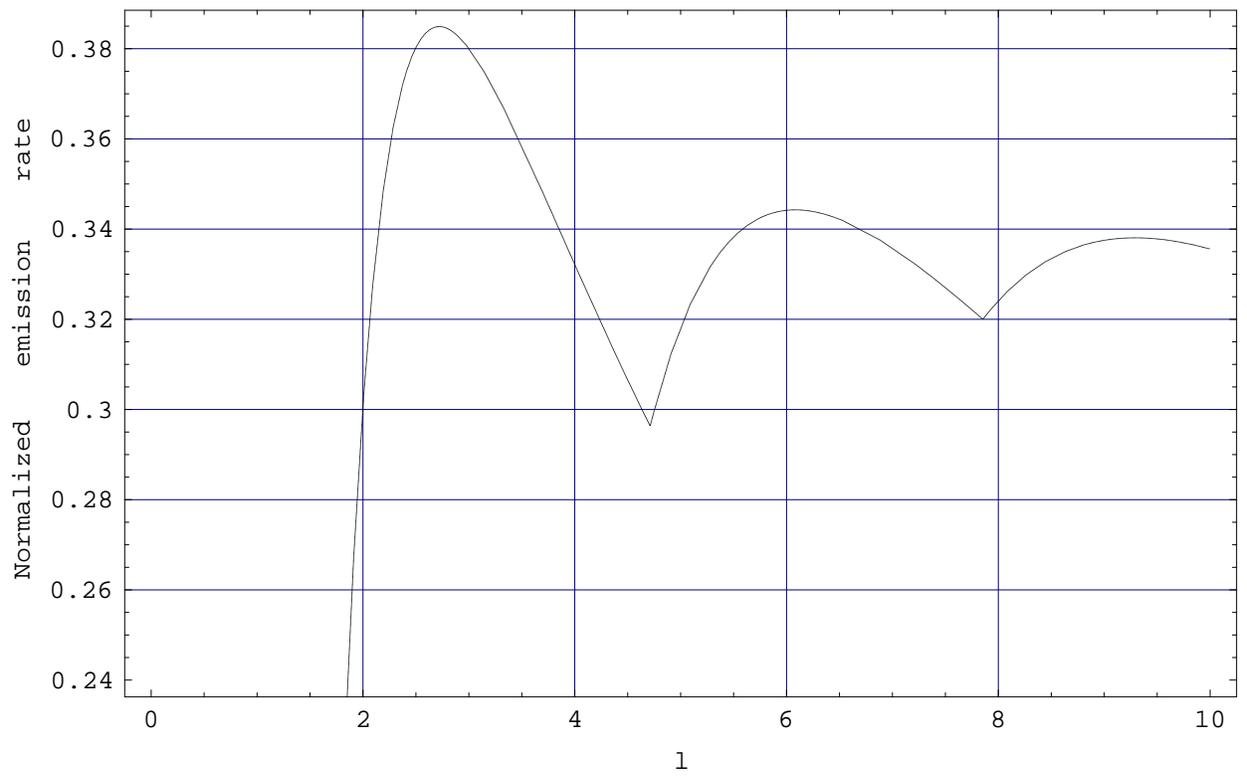

	The strong suppression that occurs in the case of two conducting plates for
$\Apa$ has its counterpart in the two cases discussed previously, as we
shall see. However, we should emphasize that in the case of two permeable
plates as well as in the case of a conductting plate and a permeable one,
the suppression occurs for $\Ape$, in contrast with the case of two
conducting plates. For simplicity, let us just fix the atom at a point
equidistant from the parallel infinite plates in both setups and vary the
distance $L$ between the plates. Also, for convenience, in the remaining
figures of this letter we shall plot the graphs of the ratios $\Apa/\Afpa$
and $\Ape/\Afpe$ as functions of the dimensionless parameter $l:=k_0L$.
Figure (\ref{3}) shows jointly the suppression of $\Ape$ for the case of two
permeable plates (solid line) and the suppression of $\Apa$ for the case of
two conducting plates (dashed line). Observe that both occur for the same
value of the distance between the plates. Though not obvious, this result is
quite reasonable, since for the case of two permeable plates the mode
functions of the vacuum field are also symmetric with respect to  $z=L/2$.
In this sense, for the case of a conducting plate and a permeable one, for
which mixed boundary conditions are used, it is natural to expect that the
suppression will occur for a different value of $L$. This is indeed what
happens and as it is shown separatly in figure (\ref{4}), the suppression
occurs for a value of $L$ which is smaller than the value found for the
other cases (shown in figure (\ref{3})).
 
%In this case suppression starts at $k_0L\approx1.57$
%
To conclude two final remarks. Firstly, it is very interesting to notice
that though the Casimir energy density for the case of two perfectly
parallel conducting plates is exactly the same as that for two infinitely
permeable parallel plates, the influence of these two different surroundings
in radiative properties of an atomic system (like the spontaneous emission
rate of an atom) can be quite different. In other words, though the Casimir
effect is \lq\lq blind" to the change of the two conducting plates by the
two infinitely permeable ones, the atom is not. The reason for that is
simply because only the possible field frequencies enter in the calculation
of the Casimir energy density, while the atom interacts directly with each
vaccum field mode, it probes locally the vacuum field. Finally, we think it
could be interesting to do experiments about the influence of the proximity
of material walls in the spontaneous emission rate of atomic systems
analogous to those mentioned before where conducting plates could be
interchanged at will with permeable ones. Comparing the results thus
obtained may add some new information to such an interesting problem as the
atom-cavity interaction.

\clearpage
\begin{figure}
\caption{The ratio  $A_{21}/A_{21}^0$ as a function of the dimensionless
variable $s=k_0z$ for the case of two perfectly conducting plates  (dashed
curve) and the case of a perfectly conducting plate and an infinitely
permeable plate (solid curve).}
\label{1}
\end{figure}
\begin{figure}
\caption{The ratio $A_{21}/A_{21}^0$ as a function of the dimensionless
variable $s=k_0z$ for the case of two perfectly conducting plates (dashed
curve) and the case of two infinitely permeable plates (solid curve).}
\label{2}
\end{figure}
\begin{figure}
\caption{The ratio  $\Apa/\Afpa$ for the case of two perfectly conducting
plates (dashed curve) and the ratio $\Ape/\Afpe$ for the case of a perfectly
conducting and infinitely permeable plates (solid curve) as functions of the
dimensionless variable $l=k_0L$.} 
\label{3}
\end{figure}
\end{document}